\documentclass[fleqn,usenatbib]{mnras}
\usepackage[T1]{fontenc}
\usepackage{ae,aecompl}
\usepackage[dvipdfmx]{graphicx}	
\usepackage{multirow}
\usepackage{amsmath}
\usepackage{amssymb}
\usepackage{ulem} 

\usepackage[dvipsnames]{xcolor}
\definecolor{color_tm}{RGB}{255,127,80}

\newcommand{\Msun}{\ensuremath{\mathrm{M}_\odot}}
\newcommand{\MP}{\ensuremath{m_{\rm{p}}}}

\title[Radio emission in AT2019dsg]
{What powers the radio emission in TDE AT2019dsg: a long-lived jet or the disruption itself? }

\author[Matsumoto et al.]{Tatsuya Matsumoto,$^{1,2,3,4}$\thanks{E-mail: tatsuya.matsumoto@mail.huji.ac.il}\thanks{JSPS Research Fellow} Tsvi Piran,$^{1}$ and Julian H. Krolik$^{5}$
\\
$^{1}$Racah Institute of Physics, Hebrew University, Jerusalem, 91904, Israel\\
$^{2}$Research Center for the Early Universe, Graduate School of Science, University of Tokyo, Tokyo 113-0033, Japan\\
$^{3}$Department of Physics, Graduate School of Science, University of Tokyo, Tokyo 113-0033, Japan\\
$^{4}$Columbia Astrophysics Laboratory, Columbia University, New York, NY 10027, USA\\
$^{5}$Physics and Astronomy Department, Johns Hopkins University, Baltimore, MD 21218, USA\\
}

\pubyear{2021}

\begin{document}
\label{firstpage}
\pagerange{\pageref{firstpage}--\pageref{lastpage}}
\maketitle

\begin{abstract}

The tidal disruption event AT2019dsg was observed from radio to X-rays and was possibly accompanied by a high-energy neutrino.
Previous interpretations  have focused on continued injection by a central engine as the source of energy for radio emission.
We show that continuous energy injection is unnecessary; the radio data can be explained by a single ejection of plasma that supplies all the energy needed.
To support this assertion, we analyze the synchrotron self-absorbed spectra in terms of the equipartition model.
Similar to previous analyses, we find that the energy in the radio-emitting region increases approximately $\propto t^{0.7}$ and the lengthscale of this region grows  $\propto t$ at a rate $\simeq 0.06\,c$. 
This event resembles the earliest stage of a supernova remnant: because the ejected mass is much greater than the shocked external mass, its velocity remains unchanged, while the energy in shocked gas grows with time.
The radio-emitting material gains energy from the outflow, not continuing energy injection by the central object.
Although energy injection from an accreting BH cannot be completely excluded, the energy injection rate is very different from the fallback luminosity, and maintaining constant outflow velocity requires fine-tuning demanding further physical explanation.
If the neutrino association is real, the energy injection needed is much greater than for the radio emission, suggesting that the detected neutrino did not arise from the radio-emitting region.
\end{abstract}


\section{Introduction}
A supermassive black hole (BH) lurking in a center of a galaxy can destroy a star approaching very close due to its strong gravity \citep{Hills1975,Rees1988}.
Tidal disruption events (TDEs) have been observed in many wavelengths: X-rays \citep{Saxton+2020}; UV/optical \citep{Roth+2020,vanVelzen+2020}; and radio \citep{Alexander+2020}.
Understanding these observations will help us to take a census of the population of supermassive BHs, revealing their mass and spin distributions.  It will also help us investigate relativistic effects around BHs by shedding light on the jet-launching mechanism, as well as stellar dynamics in galactic nuclear clusters \citep{Stone+2020}.

AT2019dsg is one of the most well-observed TDEs.
This event was discovered as a bright UV/optical TDE on 9 April 2019 by the Zwicky Transient Facility \citep{vanVelzen+2021}.
Follow-up observation revealed rapidly declining X-rays $\simeq40$ days after the discovery \citep{Cannizzaro+2021,Stein+2021} and radio emission lasting more than 500 days \citep{Stein+2021,Cendes+2021b,Mohan+2021}.
In addition, IceCube reported detection of a high-energy neutrino in the same direction as the TDE \citep{Stein+2021}, although the probability that this is actually associated with the TDE is not high.
No TDE detected since ASASSN-14li \citep{Miller+2015,Alexander+2016,Cenko+2016,Holoien+2016,Jiang+2016,vanVelzen+2016,Brown+2017} has such a rich set of multiwavelength data.

In this paper we focus on the radio, which is useful for studying the dynamics of outflow.
The radio spectra have a peak, which is caused by the synchrotron self-absorption (SSA);  applying the equipartition method to the time-dependence of the frequency of the peak and its luminosity, we can estimate both outflow properties and the density of the circumnuclear medium (CNM) surrounding the BH \citep{Chevalier1998,BarniolDuran+2013}.
\citet{Stein+2021} and \citet{Cendes+2021b} carried out such an analysis and estimated the radius and energy in the radio-emitting site as well as the CNM density, finding
that the energy in the emitting region increases with time.
Both attributed the increase to energy injection from the vicinity of the BH.  \cite{Stein+2021} rejected the possibility that the radio resulted from instantaneous ejection of outflow, while \cite{Cendes+2021b} briefly mentioned this scenario but claimed that continuous injection from the BH is more likely.
In this paper, we revisit this possibility and show that this interpretation, in which the increasing energy is caused by an instantaneously ejected outflow, is in fact the explanation of this system's behavior requiring the fewest assumptions.

We organize this paper as follows.
In \S \ref{sec method} we review the equipartition method and estimate the physical quantities of the radio-emitting outflow, correcting certain technical errors in earlier analyses.
We analyze the time evolution of the equipartition radius in \S \ref{sec radius}, the CNM density profile in \S \ref{sec CNM}, and the energy within the emitting region in \S \ref{sec:Energy}. 
In \S \ref{sec energy} we discuss the implications of the increasing energy of the radio-emitting region and find that, rather than indicating energy injection from the BH to the radio outflow, it is more naturally interpreted as the result of shock propagation in an impulsive event.
We discuss the implications of our result for possible neutrino emission and the origin of the outflow in \S \ref{sec discussion} and conclude this paper in \S \ref{sec summary}.

\section{The Equipartition method}\label{sec method}

We briefly review the equipartition method, which is applicable to a radio-emitting outflow with a spectral peak caused by the SSA \citep{Pacholczyk1970,Scott&Readhead1977,Chevalier1998,BarniolDuran+2013}.
Consider an outflow producing synchrotron emission with an SSA spectral-peak flux density $F_{\rm p}$ at a frequency $\nu_{\rm p}$.
The equipartition method gives us the radius and energy of the radio-emitting region under the assumption that the 
 energy is distributed equally to relativistic electrons and magnetic field.

\begin{table*}
\begin{center}
\caption{Radio data and results of our equipartition analysis for AT2019dsg.
The time $\Delta t$ is measured since the discovery in the observer frame (9 April 2019), which is different from that in \citealt{Cendes+2021b} who set the origin as 10 days before the discovery.
The equipartition radius, energy, Lorentz factor, and total number of emitting electrons at $\nu_{\rm p}$, and CNM density are calculated by Eqs. \eqref{eq r_eq} - \eqref{eq density}.
We adopt two (freely-coasting and decelerating)
velocity fits to the time evolution of $R_{\rm eq}$, and the corresponding density slope $k$. The spectra of the first two epochs are not of good quality and they are excluded from our analysis.}
\label{table data}
\begin{tabular}{rccccccccc}
\hline
$\Delta t$&$F_{\rm p}$&$\nu_{\rm p}$&$R_{\rm eq}$&$E_{\rm eq}$&$\gamma_{\rm e}$&$N_{\rm e}$&$n$ (freely-coasting)&$n$ (decelerating)\\
$[\rm d]$&[mJy]&[10GHz]&[$10^{16}$cm]&[$10^{48}$erg]&&[$10^{50}$]&[$10^{3}\rm cm^{-3}$]&[$10^{3}\rm cm^{-3}$]\\
\hline
(42)&($0.47\pm0.09$)&($1.58\pm0.36$)&($0.84\pm0.21$)&($0.97\pm0.44$)&($59.05\pm4.73$)&($0.68\pm0.24$)&($36.94\pm19.67$)&($34.20\pm18.21$)\\
(45)&($0.60\pm0.04$)&($2.09\pm0.63$)&($0.72\pm0.22$)&($0.99\pm0.43$)&($59.84\pm4.78$)&($0.67\pm0.23$)&($60.87\pm39.87$)&($58.60\pm38.38$)\\
50&$0.67\pm0.01$&$1.82\pm0.13$&$0.87\pm0.07$&$1.29\pm0.41$&$60.20\pm4.81$&$0.87\pm0.15$&$45.08\pm13.50$&$45.97\pm13.77$\\
70&$0.80\pm0.06$&$1.38\pm0.19$&$1.25\pm0.18$&$2.11\pm0.75$&$60.79\pm4.87$&$1.40\pm0.32$&$24.96\pm9.59$&$30.25\pm11.62$\\
72&$0.65\pm0.03$&$1.07\pm0.07$&$1.46\pm0.12$&$2.12\pm0.69$&$60.10\pm4.80$&$1.43\pm0.26$&$15.74\pm4.71$&$19.34\pm5.79$\\
120&$1.24\pm0.05$&$1.02\pm0.09$&$2.07\pm0.21$&$4.81\pm1.60$&$62.25\pm5.02$&$3.07\pm0.59$&$12.47\pm4.06$&$19.48\pm6.35$\\
151&$0.98\pm0.04$&$0.95\pm0.09$&$1.98\pm0.20$&$3.89\pm1.29$&$61.46\pm4.94$&$2.53\pm0.48$&$11.43\pm3.71$&$19.79\pm6.43$\\
178&$1.22\pm0.04$&$0.51\pm0.05$&$4.09\pm0.42$&$9.42\pm3.13$&$62.20\pm5.01$&$6.01\pm1.14$&$3.14\pm1.02$&$5.85\pm1.91$\\
290&$0.79\pm0.04$&$0.35\pm0.04$&$4.82\pm0.60$&$8.08\pm2.74$&$60.75\pm4.87$&$5.37\pm1.10$&$1.65\pm0.58$&$3.80\pm1.34$\\
551&$0.34\pm0.04$&$0.17\pm0.04$&$6.76\pm1.77$&$6.13\pm2.59$&$58.02\pm4.60$&$4.40\pm1.43$&$0.46\pm0.26$&$1.37\pm0.78$\\
\hline
\end{tabular}
\end{center}
\end{table*}

We follow the expressions for the radius and energy of \cite{BarniolDuran+2013}, but with some minor corrections.\footnote{We take into account an additional dependence on radius $R$ in the relativistic electrons' energy, $\propto R^{2(1-p)}$, which is ignored by \cite{BarniolDuran+2013}. This adds minor correction factors in their original expressions.}
We assume that an arbitrary fraction of the dissipated kinetic energy is transferred to the electrons $\varepsilon_{\rm e}$ and magnetic field $\varepsilon_{\rm B}$ (see their equations 27 and 28 in the Newtonian limit):
\begin{align}
&R_{\rm eq}\simeq1\times10^{17}{\,\rm cm\,}\biggl[21.8\times(525)^{p-1}\gamma_{\rm m}^{2-p}\biggl(\frac{p+1}{3}\biggl)\biggl]^{\frac{1}{2p+13}}
	\label{eq r_eq}\\
&\Big[F_{\rm p,mJy}^{\frac{p+6}{2p+13}}d_{\rm L,28}^{\frac{2(p+6)}{2p+13}}\nu_{\rm p,10}^{-1}(1+z)^{-\frac{3p+19}{2p+13}}\Big]
f_{\rm A}^{-\frac{p+5}{2p+13}}f_{\rm V}^{-\frac{1}{2p+13}}
\Big\{(4\xi\epsilon)^{\frac{1}{2p+13}} \Big\}\ ,
	\nonumber
\end{align}
and 
\begin{align}	
&E_{\rm eq}\simeq1.3\times10^{48}{\,\rm erg\,}\biggl(\frac{2p+13}{17}\biggl)\biggl[21.8\biggl(\frac{p+1}{3}\biggl)\biggl]^{-\frac{2(p+1)}{2p+13}}
	\label{eq e_eq}\\
&\big[(525)^{p-1}\gamma_{\rm m}^{2-p}\big]^{\frac{11}{2p+13}}\Big[F_{\rm p,mJy}^{\frac{3p+14}{2p+13}}d_{\rm L,28}^{\frac{2(3p+14)}{2p+13}}\nu_{\rm p,10}^{-1}(1+z)^{-\frac{5p+27}{2p+13}}\Big]
	\nonumber\\	
& f_{\rm A}^{-\frac{3(p+1)}{2p+13}}f_{\rm V}^{\frac{2(p+1)}{2p+13}}\biggl\{(4\xi)^{\frac{11}{2p+13}}\biggl[\frac{11}{2p+13}\epsilon^{-\frac{2(p+1)}{2p+13}}+\frac{2(p+1)}{2p+13}\epsilon^{\frac{11}{2p+13}}\biggl] \biggl\}
	\nonumber \ .
\end{align}
Here the observables are the power-law index of the electron distribution $p$ (given by the spectral slope) and the luminosity distance corresponding to the redshift $z$, $d_{\rm L}$, as well as $F_{\rm p,mJy}(=F_{\rm p}/\rm mJy)$ and $\nu_{\rm p}$.
We use the notation $Q_x=Q/10^x$ in cgs units unless otherwise specified.
$\gamma_{\rm m}$ is the minimal Lorentz factor of relativistic electrons.
In the standard prescription to estimate $\gamma_{\rm m}$ \citep[e.g.][]{Sari+1998}, it becomes less than unity for a non-relativistic outflow with a velocity $v\lesssim v_{\rm DN}\simeq0.1\,c\,[(p-2)\varepsilon_{\rm e,-1}/(p-1)]^{-1/2}$ \citep[the so-called deep-Newtonian phase,][]{Huang&Cheng2003,Sironi&Giannios2013}.\footnote{When the outflow expands at larger velocity than $v\gtrsim v_{\rm DN}$, $\gamma_{\rm m}$ depends on the outflow velocity and we have to determine $R_{\rm eq}$ and $v$ at the same time.}
In this phase, we set $\gamma_{\rm m}=2$ because only relativistic electrons produce synchrotron emission.
The area and volume-filling fractions of the outflow are defined so that the emission area and volume are given by $f_{\rm A}4\pi R^2$ and $f_{\rm V}\pi R^3$, respectively.
We also define  $\epsilon\equiv\big[(\varepsilon_{\rm B}/\varepsilon_{\rm e})/(2(p+1)/11)\big]$ and $\xi\equiv1+\varepsilon_{\rm e}^{-1}$ (see below for their meanings).   

The term ``equipartition energy'' is often used just for the sum of equally-distributed energy of the relativistic electrons and the accompanying magnetic field.
However, the actual energy could be larger, and this possible increase is expressed by the terms in curly brackets in Eqs. \eqref{eq r_eq} and \eqref{eq e_eq} (besides a factor of 4 arising from a correction to the isotropic number of radiating electrons in the non-relativistic case).
The factor $\epsilon$ describes a deviation of the ratio of magnetic to electron energy from $2(p+1)/11$, the value for the minimal energy case.
The factor $\xi$ represents additional energy stored in the (hot) protons of the emitting region.
With these corrections, the equipartition energy in Eq.~\eqref{eq e_eq} is the sum of the energy of (hot) protons, relativistic electrons, and magnetic field in the emitting site.
Note that the outflow's total energy (including kinetic energy, for example) could be much larger, as we discuss later. 

Importantly the factors $\xi$ and $\epsilon$ as well as the geometrical parameters $f_{\rm A}$ and $f_{\rm V}$ hardly influence $R_{\rm eq}$, so this argument enables us to determine the emitting radius with only small uncertainty.
The most significant dependence of $R_{\rm eq}$ comes from the covering fraction of the emitting region, $f_{\rm A}$, and it is roughly given by $R_{\rm eq} \propto f_{\rm A}^{-0.5}$.

The Lorentz factor and number of radiating electrons at $\nu_{\rm p}$ are given by \citep[equations 14 and 15 in][]{BarniolDuran+2013}
\begin{align}
\gamma_{\rm e}&\simeq525\,F_{\rm p,mJy}d_{\rm L,28}^2\nu_{\rm p,10}^{-2}(1+z)^{-3}f_{\rm A}^{-1}R_{\rm eq,17}^{-2} \ ,\\
N_{\rm e}&\simeq4.0\times10^{54}\,F_{\rm p,mJy}^3d_{\rm L,28}^6\nu_{\rm p,10}^{-5}(1+z)^{-8}f_{\rm A}^{-2}R_{\rm eq,17}^{-4} \ ,
    \label{eq number}
\end{align}
respectively.

We can now relate $N_{\rm e}$ to the external density if we assume that the emitting electrons originate from shock-heated CNM.
In this case the (pre-shock) CNM density is estimated by dividing the total number of electrons by the volume swept up by the outflow:
\begin{align}
n=&\frac{N_{\rm e}\big(\frac{\gamma_{\rm m}}{\gamma_{\rm e}}\big)^{1-p}{\rm max}\big[\big(\frac{v_{\rm DN}}{v}\big)^2,\,1\big]}{\frac{\Omega}{3-k}R_{\rm eq}^3} \simeq4100{\,\rm cm^{-3}\,}
    \label{eq density}\\
&\times \big[525\gamma_{\rm m}^{-1}\big]^{p-1}F_{\rm p,mJy}^{p+2}d_{\rm L,28}^{2(p+2)}\nu_{\rm p,10}^{-2p-3}(1+z)^{-3p-5}
	\nonumber\\
&\times f_{\rm A}^{-p-1}R_{\rm eq,17}^{-2p-5}\Omega^{-1}(3-k){\rm max}\biggl[\Big(\frac{v_{\rm DN}}{v}\Big)^2,\,1\biggl]\ .
	\nonumber
\end{align}
Here three correction factors arise. First, most relativistic electrons have Lorentz factor $\sim \gamma_{\rm m}$, and their number is $(\gamma_{\rm m}/\gamma_{\rm e})^{1-p}$ times larger than $N_{\rm e}$.
Second, when the outflow is in the deep-Newtonian phase, only a fraction $(v/v_{\rm DN})^2$ of electrons are accelerated to relativistic energy \citep{Sironi&Giannios2013,Matsumoto&Piran2021b}.\footnote{Neglecting this factor means the fraction of energy transferred to relativistic electrons is larger than $\varepsilon_{\rm e}$.}
Third, the density is assumed to follow $n\propto R^{-k}$ ($k<3$), and the total electron number in the volume swept-up by the outflow with a constant solid angle $\Omega$ is given by $\int_0^R \Omega r^2 n(r)dr=\Omega n (R)R^3/(3-k)$.
The index $k$ is known only after the density profile is calculated without this factor.
When the density profile is not described by a power-law function, we can approximate the integral as $\simeq\Omega n(R) R^3$, which is justified as long as $d\log n/d\log r<3$.

We apply the equipartition method to AT2019dsg using the peak fluxes and frequencies in Table \ref{table data} taken from \cite{Stein+2021} and \cite{Cendes+2021b}.
The other observables are $z=0.051$, the corresponding luminosity distance $d_{\rm L}\simeq230\,\rm Mpc$, and the power-law index $p=2.7\pm0.2$.
We adopt the same parameters as those of \cite{Cendes+2021b}, who assumed that the outflow is spherical ($\Omega=4\pi$), and the emitting region is a shell with the width of $dR=0.1\,R_{\rm eq}$. These assumptions give $f_{\rm A}=1$ and $f_{\rm V}=0.36$.
The other parameters are $\varepsilon_{\rm e}=0.1$, $\varepsilon_{\rm B}=0.02$, and $\gamma_{\rm m}=2$ (assuming $v<v_{\rm DN}$, which is justified later).
The results are shown in Table \ref{table data}, which are basically consistent with those of \cite{Cendes+2021b} except for the CNM density; in that estimate, they neglected the correction factor $(v/v_{\rm DN})^2$.
For the outflow velocity, we use two (freely-coasting and decelerating) fits to the time evolution of $R_{\rm eq}$ (see the next section).
We also adopt the density slope $k$ obtained by fitting the calculated profile.

\section{Results of the equipartition analysis}
\subsection{Launch time of the radio-outflow}\label{sec radius}

\begin{figure}
\begin{center}
\includegraphics[width=85mm, angle=0]{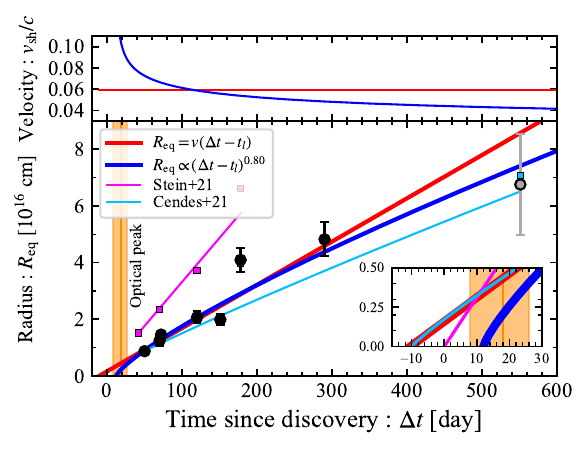}
\caption{({\bf Bottom}) Equipartition radius at each observation epoch.
The red and blue curves denote the best-fit curves of freely-coasting and decelerating fits, respectively, with key parameters of $v_{\rm sh}\simeq0.059\,c$ and $t_l\simeq-10\,\rm d$ (freely-coasting) and $\alpha=0.80$ and $t_l=12\,\rm d$ (decelerating).
The gray dot highlights the lower quality of the spectrum at 551 days.
Also shown are the results by \citealt{Stein+2021} (magenta squares) and \citealt{Cendes+2021b} (light-blue squares).
\citealt{Stein+2021} assumed a conical outflow, which results in larger radii.
The orange shaded region shows the peak time of optical emission.
The inset depicts the best-fit curves around the time of discovery defined as $\Delta t=0$.
({\bf Top}) Outflow velocity derived by time derivative of the best-fits of $R_{\rm eq}$.}
\label{fig radius}
\end{center}
\end{figure}

To fit the time evolution of the equipartition radius, we derive the velocity and launching time of the radio-emitting outflow (hereafter  radio-outflow). 
Fig. \ref{fig radius} depicts the equipartition radius at each observation time.
Also shown in this figure are the results of \cite{Stein+2021} and \cite{Cendes+2021b}.
Following \cite{Stein+2021} we use the discovery date as the origin of time.
Note that \cite{Cendes+2021b} set their origin of time 10 days earlier.
Since the quality of the data at 42 and 45 days after the discovery is too poor to determine the spectral peak, we do not include these observations in the following calculation.
The frequency of the peak is also somewhat difficult to determine in the data at 551 days, but we include it in analysis with a sizable error bar.

\begin{figure}
\begin{center}
\includegraphics[width=85mm, angle=0]{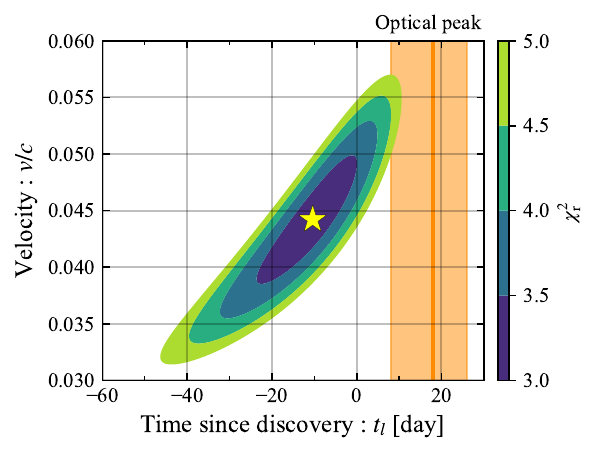}
\caption{Distribution of the reduced chi-square for the freely-coasting fit of $R_{\rm eq}=v(\Delta t-t_l)$.
The yellow star denotes the best-fit parameter set of $v=v_{\rm sh}/\zeta\simeq0.04\,c\simeq13000\rm \,km\,s^{-1}$ and $t_l\simeq-10\,\rm d$.
}
\label{fig chi}
\end{center}
\begin{center}
\includegraphics[width=85mm, angle=0]{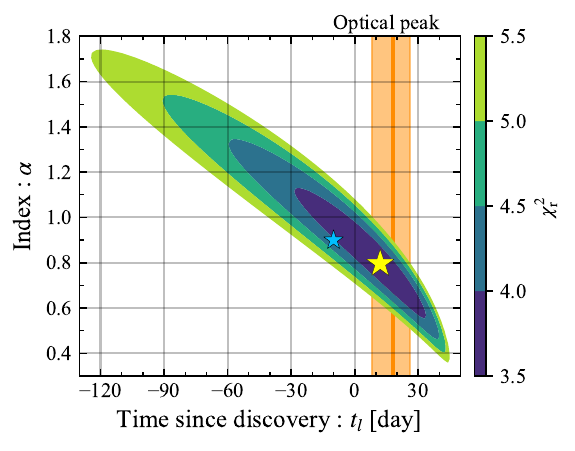}
\caption{The same as Fig. \ref{fig chi} but for the deceleration fit of $R_{\rm eq}\propto (\Delta t-t_l)^{\alpha}$.
The best-fit parameters are $\alpha\simeq0.80$ and $t_l\simeq12\,\rm d$.
{The light-blue star denotes the location of the fit by \citealt{Cendes+2021b}.}
}
\label{fig chi2}
\end{center}
\end{figure}

We consider two possible fits to the data.
One, which we call constant velocity, corresponds to freely-coasting outflow: $R_{\rm eq}=v_{\rm sh}(\Delta t-t_l)$, where $\Delta t$ and $t_l$ are the observation time and outflow-launching time, respectively (both are measured since the discovery).
For our chosen  equipartition parameters, the best-fit parameters are $v_{\rm sh}=0.059_{-0.005}^{+0.005}\,c\simeq18000_{-1500}^{+1500}\,\rm km\,s^{-1}$ and $t_l=-10_{-9}^{+7}\,\rm d$; that is, the outflow begins before the optical detection (the error represents $1\sigma$).  For this fit, the reduced chi-square is $\chi_{\rm r}^2\simeq3.1$.
The second fit is a power-law that corresponds to a decelerating (or accelerating) outflow with $R_{\rm eq}=A(\Delta t-t_l)^\alpha$.
The best-fit parameters are $t_l=12_{-26}^{+17}\,\rm d$ and
$\alpha=0.80_{-0.18}^{+0.22}$.
These parameters imply an outflow that is launched around the time of the optical detection,
and whose speed decelerates very gradually.
For this fit, the reduced $\chi_{\rm r}^2\simeq3.6$.
Note that the estimate of the  launching time $t_l$ is independent of the equipartition parameters $\epsilon$, $\xi$, $f_{\rm A}$, and $f_{\rm V}$ \citep{Krolik+2016}. 
Figs. \ref{fig chi} and \ref{fig chi2} depict the distribution of reduced $\chi_{\rm r}^2$ for two fits.
The quality of fit in both cases is similar, and we cannot favor one fit from the other.
However, the the evidence for deceleration depends entirely on the $\Delta t=551$~d observation, which has a large uncertainty.

\cite{Stein+2021} assumed a conical outflow ($f_{\rm A}=0.13$, $f_{\rm V}=1.15$, $\varepsilon_{\rm e}=0.1$, and $\varepsilon_{\rm B}=10^{-3}$) and derived radii somewhat larger than ours because of the smaller opening angle.
They fitted the data up to 178 days and found the best fit for a constant velocity $v_{\rm sh}=0.12\,c$.
\cite{Cendes+2021b} found that the evolution is best fitted by a slowly decelerating fit with $\alpha=0.9$, i.e., slightly slower deceleration than in our power-law fit.
The parameters we adopted in our equipartition analysis are almost identical with theirs.
However, they fixed the launching time at $10\, \rm days$ before the discovery by extrapolating the optical light curve to zero flux and assuming that was the launch time, while we infer it from a fit to $R_{\rm eq}$. 
For our freely-coasting fit, our fit leads to the same launch time as assumed by \citet{Cendes+2021b}, while for the decelerating model the best-fit value $\simeq10\,\rm d$ after the discovery, but with a large uncertainty (see Fig.~\ref{fig chi2}). As can be seen in Fig. \ref{fig chi2}, the result of \citet{Cendes+2021b} is within the error of both of our estimates.

The top panel of Fig. \ref{fig radius} depicts the velocity of each model as a function of time, deriving it from $v_{\rm sh}= dR_{\rm eq}/d(\Delta t)$.
This quantity represents  the pattern velocity of the emitting region. If the emission is due to a shock propagating in the CNM then this is the speed of the shock front. In this case the actual outflow velocity, $v$, is somewhat smaller than the shock velocity; their ratio is given by the shock jump condition \citep[e.g.][]{Landau&Lifshitz1987}: $\zeta\equiv v_{\rm sh}/v=(\hat{\gamma}+1)/2$, where $\hat{\gamma}$ is the adiabatic index of the shocked material.
For $\hat{\gamma}=5/3$, $\zeta=4/3$. This factor is order unity but it  is important for inferring the origin of the outflow.

\subsection{The CNM density profile}\label{sec CNM}
\begin{figure}
\begin{center}
\includegraphics[width=85mm, angle=0]{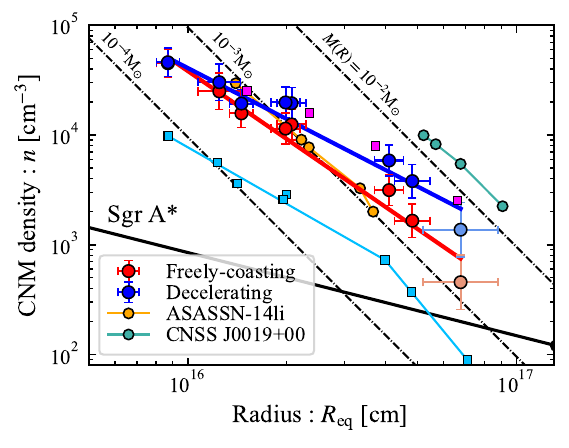}
\caption{CNM density profiles reconstructed from the equipartition analysis.
Red and blue line shows the best-fit power-law functions for the freely-coasting ($n\propto R^{-2.1}$) and decelerating ($n\propto R^{-1.6}$) fits for AT2019dsg, respectively.
Magenta and light-blue squares show the results by \citealt{Stein+2021} (almost overlapping our decelerating fit) and \citealt{Cendes+2021b}, respectively.
Profiles of Milky Way \citep[$n \propto R^{-1}$,][]{Baganoff+2003,Gillessen+2019} and other TDEs  ($n\propto R^{-2.5}$) are also presented.
Black dash-dotted lines show the positions where the enclosed mass of $M(R)\simeq4\pi \MP n R^3=10^{-4}-10^{-2}\,\Msun$.
}
\label{fig profile}
\end{center}
\end{figure}

Radio observations of TDEs provide possibly the only way to infer the CNM density distribution around distant galaxies \citep{Krolik+2016,Alexander+2016}.  
Fig. \ref{fig profile} depicts the profiles reconstructed by our equipartition analysis.
The two thick colored curves represent the predictions made by our two outflow models (note that the estimate of the CNM density depends on the outflow velocity, see Eq.~\ref{eq density}).
The velocities are derived from the best-fits of the equipartition radius, $v=\zeta^{-1} dR_{\rm eq}/d(\Delta t)$.
Fitting the density profiles with a power-law function $n\propto R^{-k}$, we find the best fits for $k=2.06_{-0.16}^{+0.17}$ and $1.55_{-0.16}^{+0.17}$ for the freely-coasting and decelerating fits, respectively.

Fig. \ref{fig profile} also depicts the density profiles obtained by \cite{Stein+2021} and \cite{Cendes+2021b}.
\cite{Stein+2021} derived the density of non-thermal electrons as a lower-limit of the CNM density.
Lacking a detailed  description of  their derivation  we cannot compare  their results to ours. 
\cite{Cendes+2021b} obtained a density profile that is about 5 times smaller than ours.
As \cite{Cendes+2021b} neglected  the deep-Newtonian correction factor  their results should have been $\simeq14$ times smaller than ours (with $v/v_{\rm DN} \simeq0.27$).
However, at the same time they divided the total number of electrons by the emitting volume, $\pi f_{\rm V}R^3$, and by the shock compression factor 4, instead of dividing it by the volume swept up by the shock.
This somewhat compensates the difference, resulting in a density profile 5 times lower than ours.

The CNM profiles of our Galactic center and the host galaxies of two other radio TDEs, ASASSN-14li \citep{Alexander+2016,Krolik+2016} and CNSS J0019+00 \citep{Anderson+2020} are also shown in Fig. \ref{fig profile}.
The latter two are derived by the same procedure that we used for AT2019dsg.
Their equipartition radii evolve at a constant velocity of $v_{\rm sh}\simeq0.04\,c$ and the deep-Newtonian correction becomes more important for them.
The CNM densities in all the TDE host galaxies are larger than in Sgr A*.
The CNM density profile of AT2019dsg is shallower than in the other TDEs; this contrast is related to the different time evolution of its equipartition energy.

\subsection{The Energy of the emitting region}
\label{sec:Energy}

Using the equipartition analysis, we can estimate two different energies associated with the emitting region.
First, using Eq.~\eqref{eq e_eq} we obtain the equipartition energy, which is the sum of the internal energies of the protons and electrons plus the magnetic field energy within the emitting region.

\begin{figure}
\begin{center}
\includegraphics[width=85mm, angle=0]{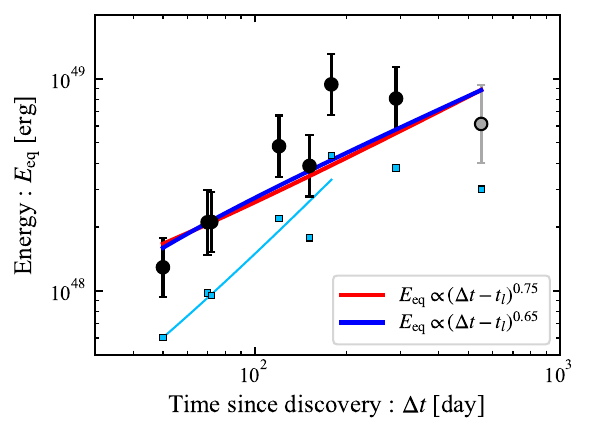}
\caption{The equipartition energy at each observation epoch.
The red and blue curves represent the best-fit power-law functions with the same outflow-launching time as the freely-coasting ($t_l=-10\,\rm d$) and decelerating  ($12\,\rm d$) outflow fits, respectively. 
Light-blue squares denote the result by \citealt{Cendes+2021b}.
}
\label{fig energy}
\end{center}
\end{figure}

In Fig. \ref{fig energy} we show $E_{\rm eq}$ as we compute it for AT2019dsg and contrast it with the values found by \citet{Cendes+2021b}. We also fit the time evolution of $E_{\rm eq}$ by a power-law function 
$E_{\rm eq}\propto t^{\beta}$, 
where $t\equiv \Delta t-t_l$,
using $t_l$ obtained from the fits to $R_{\rm eq}$.
We find that  $\beta=0.75_{-0.15}^{+0.14}$ for the freely-coasting fit and $\beta=0.65_{-0.13}^{+0.12}$ for the deceleration fit.
\cite{Stein+2021} and \cite{Cendes+2021b} also found that
the energy rises over time: $E_{\rm eq}\propto \tilde{t}$ and $\tilde{t}^{1.5}$, respectively, where $\tilde{t}$ is the time measured since their time origins.
Both papers found a faster rate of energy increase than we do because they used data points only up to 178 days.  More importantly, all estimates find that the energy within the radio-producing region increases as time goes on, although there is a hint of a possible saturation at late times.

Combining the mass of the radio-emitting region (Eq.~\ref{eq density}) with the speed, $v$, gives an estimate of the emitting region's kinetic energy : 
\begin{align}
E_{\rm kin}= \frac{f_x}{2}Mv^2 = f_x 2\pi\MP n R^3 v^2 \propto {t}^{(5-k)\alpha -2}\ ,
\label{eq:Ek}
\end{align}
where $f_x$ is a numerical factor of order unity, and we used $R\propto t^{\alpha}$ and $n\propto R^{-k}$ in the last expression.
For the values of $\alpha$ and $k$ we found in \S~\ref{sec radius} and \S~\ref{sec CNM}, the time exponent in Eq.~\eqref{eq:Ek} becomes $\simeq0.94_{-0.17}^{+0.16}$ and $0.76_{-0.60}^{+0.77}$
for the freely-coasting and decelerating fits, respectively.  Both inferred exponents are consistent with our estimates for $\beta$ to within their uncertainties. The large error bars in the fits are dominated by the last data point ($\Delta t=551\,\rm d$) in the analysis.\footnote{Ignoring the last data point we obtain for the freely-coasting fit $k=1.8$, $\beta=1.2$ (showing only the central values) and consistently $(5-k)\alpha-2=1.2$, and for the decelerating fit $\alpha=0.96$, $k=1.8$,  $\beta=1.1$, and  $(5-k)\alpha-2=1.1$.}

Eq.~\eqref{eq:Ek} can be rewritten in terms of the parameters inferred through equipartition analysis (Eqs. \ref{eq r_eq} and \ref{eq e_eq} and the velocity, $v$, inferred from  $dR_{\rm eq}/dt$).
{Usually} this provides an independent check on the energy that can be compared with $E_{\rm eq}$.
However, applying the deep Newtonian correction leads to a cancellation of the $v^2$ factor and a dependence on the observable quantities $F_p$ and $\nu_p$ identical to that of $E_{\rm eq}$.
As a result, $E_{\rm kin}$ and $E_{\rm eq}$ are not independent; their ratio $E_{\rm kin}/E_{\rm eq}$ is constant and $\sim O(1)$.

\section{Instantaneously ejected outflow or Continuous Energy Injection?}\label{sec energy}

We can now address the question raised in the Introduction: is energy continually injected into the outflow from the central object, as inferred by \cite{Stein+2021} and \cite{Cendes+2021b}, or is the radio emission powered by an outflow ejected instantaneously, possibly during the tidal disruption process? Put differently, is there an active  central engine that powers the radio emission or are we observing the interaction of the tidal debris with the CNM?

\cite{Stein+2021} and \cite{Cendes+2021b} suggested that the energy growth in the radio-emitting region is caused by a continuous energy injection to the shocked region from the accreting BH.
Both papers discarded the possibility that it could be caused by interaction with an outflow ejected instantaneously and having a spread of velocity, using different erroneous arguments.
\cite{Stein+2021}, who analyzed the data only up to 178 days, rejected this idea because their last point gave a somewhat faster velocity, arguing that outflow cannot accelerate.
Although this is true, such an outflow cannot accelerate, too much weight was given in this argument to a single data point: later data points show no acceleration and the whole data set is best fit with either a constant velocity or a mild deceleration model.
\cite{Cendes+2021b} mentioned instantaneous ejection with a spread of velocities, but preferred continuous injection because the velocity is roughly constant.
In the following, we show that this second argument is also incorrect.

\subsection{Instantaneously ejected outflow}
When there is an instantaneously ejected outflow, the first stage of the interaction between the outflow   and the surrounding matter, like in a supernova remnant, involves a combination of two shocks: a forward shock propagating into the CNM and a reverse shock propagating into the outflow.  These post-shock regions are separated by a contact discontinuity. As long as the mass of the outflow is significantly larger than the shocked CNM mass, the system effectively does not slow down.  Because the forward shock moves at a speed similar to that of the unshocked ejecta, the reverse shock velocity is much smaller than the forward shock velocity, making it a much weaker shock.
Consequently, the contribution of the reverse-shock heated material to the overall luminosity is negligible.

As more CNM is swept up and heated by the forward shock, more energy is transferred to the shocked region at the expense of the very large kinetic energy of the outflow.\footnote{Energy loss by electron synchrotron cooling is negligible because the cooling timescale is much longer than the dynamical time, $t_{\rm cool}\simeq4500{\,\rm d\,}(B/{\rm G})^{-2}(\gamma_{\rm m}/2)^{-1}$, where the magnetic field is estimated by the equipartition analysis, $B\simeq0.1-1\,\rm G$ \citep[see also][]{Cendes+2021b}.}
Some of this energy transfer appears as an increase of the energy of the relativistic electrons and magnetic field, and hence as an increase of the equipartition energy; the rest goes into kinetic energy of the shocked gas.

In a realistic blast wave, there is some range of speeds within the ejecta.
The ejected mass therefore expands homologously with the fastest material farthest out. At any moment the energy of the emitting region is (up to a factor of order unity) the kinetic energy of the fraction of the outflow that was moving with a velocity larger than the current velocity, $E_{\rm kin}(>v)$. Eq.~\eqref{eq:Ek} implies now that 
\begin{equation}
E_{\rm kin}(>v)   \propto v^{-\frac{(5-k)\alpha -2}{ 1-\alpha}} \ . 
\label{eq:ene2}
\end{equation}
As $\alpha \simeq 1$ for both fits we see that the power in the exponent, $[(5-k)\alpha -2]/(\alpha-1)$, is very large and negative.
Thus, any outflow with a steep energy vs. velocity distribution will naturally produce the observations. Determination of the exact slope requires in this case a precise determination of $\alpha$, but given the deceleration fit we can say that the slope of $E_{\rm kin}(>v)$ should be steeper than $-4$, a reasonable value.

Thus, the observation of almost constant expansion velocity for the emitting region is a prediction of this scenario provided that the total ejected kinetic energy is significantly larger than the energy of the radio-emitting plasma, and the observed velocity is at the high-velocity tail of the velocity distribution.
In fact, this prediction has already been made in the context of other TDEs, ASASSN-14li and CNSS J0019+00, for which $dR_{\rm eq}/dt$ was likewise constant at a speed $\simeq 15000\,\rm km\,s^{-1}$ over a span of many months \citep{Krolik+2016,Anderson+2020}.
These TDEs show  $E_{\rm eq}\propto t^{0.5}$ \citep{Alexander+2016,Anderson+2020}, consistent with the expected evolution from Eq.~\eqref{eq:Ek}  for $\alpha=1$ and a slightly steeper density profile ($k=2.5$).

For AT2019dsg, a total kinetic energy  $\gtrsim 10^{49}$~erg is required to power the radio emission. This can be  easily satisfied if the outflow is the unbound debris ejected at the time of disruption, whose expected energy is $\sim 2 \times 10^{50} (M_*/\Msun)\,\rm erg$, where $M_*$ is the disrupted stellar mass (see further discussion of this scenario in \S \ref{Sec:what}).  
An interesting corollary of this idea is that the 
the outflow velocity associated with the inferred shock speed is $\simeq 13000$~km~s$^{-1}$ (see the discussion at the end of \S~\ref{sec radius}).
As we have previously emphasized, the material driving the shock belongs to the small fraction of the ejecta with the greatest speed, so such an outflow energy and speed are consistent with theoretical expectations \citep{Ryu+2020}.

\subsection{Energy injection}
\label{sec:energy-injection}

In the energy injection model, the energy for the radiating electrons is supplied by a jet possibly emerging from an accretion disk surrounding the BH. First we note that consistency with the steadily increasing $E_{\rm eq}$ requires an energy injection rate
\begin{align}
L_{\rm in}\sim\frac{dE_{\rm eq}}{dt}\simeq3\times10^{41}{\,\rm erg\,s^{-1}\,}
\biggl(\frac{t}{38{\,\rm d}}\biggl)^{-0.35}\ ,
    \label{eq luminosity}
\end{align}
where we used the parameters of the decelerating fit and the normalization for a spherical geometry.\footnote{$L_{\rm in}$ scales as $ f_{\rm A}^{-0.6}f_{\rm V}^{0.4}$ for  different geometries (see Eq.~\ref{eq e_eq}). Since both $f_{\rm A}$ and  $f_{\rm V}$ decrease for a jet the overall dependence on the geometry is weak.}
This injection luminosity is much smaller than the fallback luminosity onto the BH, $\dot{M}_{\rm fb}c^2\sim10^{46}{\,\rm erg\,s^{-1}\,}(t/38{\,\rm d})^{-5/3}R_{*,0}M_{*,0}^{1/3}M_{\rm BH,6}^{-2/3}$ \citep{Rees1988,Phinney1989}, where $R_*$, $M_*$, and $M_{\rm BH}$ are the disrupted stellar radius and mass, and the BH mass, respectively, normalized to solar values.  It also decays much slower than the putative fallback luminosity ($\propto t^{-0.35}$, rather than $\propto t^{-5/3}$).
We cannot exclude this possibility, but we do not see a simple physical explanation for  why it should be such a small fraction of the accretion power or why that fraction should grow with time.  This luminosity is also much smaller than the injection rate inferred from the neutrino association \citep[see below]{Stein+2021,Winter&Lunardini2021,Wu+2021}.

Second, the velocity of the emitting region is $dR_{\rm eq}/dt$ no matter whether a forward or a reverse shock (see below) produces the radio, and it is almost constant.  Although possible, this  requires some sort of coordination between the activity of the central engine, which supplies the energy, and the external mass distribution, the other quantity determining  this velocity.

Third, energy injection by a jet produces both a forward and  a reverse shock. Generally, the velocity of both shocked regions (forward and reverse) relative to the observer frame  is comparable to  $dR_{\rm eq}/dt$.
To catch up with the emitting region, the energizing jet must move with a velocity, denoted by $v_{\rm j} $, larger than $dR_{\rm eq}/dt$.
If $v_{\rm j} \gg dR_{\rm eq}/dt$, as would be the case, for example, if the jet is relativistic, the reverse shock would be much stronger than the forward one. If the observed radio  is produced by the forward shock, this would imply a much stronger emission from the reverse shock in some other wavelength, but this is not observed.
Alternatively, if the reverse shock provides the energy for the radio emission, the internal energy of the gas suffering the reverse shock must be comparable to the kinetic energy of the upstream material in the shock frame. If $v_{\rm j} \gg dR_{\rm eq}/dt$, this internal energy, which is $E_{\rm eq}$, will be $\gg E_{\rm kin}$, the kinetic energy of the shocked region as measured in the observer frame.  However, the observation-based equipartition analysis finds that $E_{\rm eq} \sim E_{\rm kin}$.

We must therefore conclude that $v_{\rm j} \gtrsim  dR_{\rm eq}/dt$. Clearly,  this rules out a relativistic jet. Moreover, this requires more fine-tuning, as a priori there is no relation between the two velocities.  A jet speed similar to the radio-emitting region expansion speed also implies that the jet emerges from the source more or less at the same time as the original outflow (e.g., the moment of the disruption if the outflow is unbound debris, see below). Both results are unlikely within the injection scenario.

\section{Discussion}\label{sec discussion}

\subsection{What launches the outflow?}
\label{Sec:what}

Turning now to the origin of the radio outflow, there are several possible channels that might launch Newtonian outflows from TDEs.
One is ejection of unbound debris in the course of the disruption \citep{Strubbe&Quataert2009,Krolik+2016,Yalinewich+2019b}.
Alternatively, a disk wind may be launched when debris fallback at a super Eddington rate forms an accretion disk \citep{Strubbe&Quataert2009,Dai+2018}. 
It has also been suggested that debris stream self-intersections can lead to outflows \citep{Lu&Bonnerot2020}, but even in a global simulation with exceptionally strong self-intersections \citep{Sadowski+2016}, the outflow was at most comparable, in terms of mass and specific energy, to the unbound ejecta that would be released in the disruption.
Because all these outflows are launched at different times, it may be possible to discriminate between them on the basis of our inferred launching times, $\simeq30\,\rm d$ before and at the same time as the optical peak, for the freely-coasting and decelerating fits, respectively.

These launch times, which are independent of the equipartition parameters, can constrain the outflow's origin when placed in the context of an optical emission model. 
According to the apocenter-shock model \cite[e.g.][]{Piran+2015}, the returning stream passes the pericenter having suffered little dissipation and collides with later-arriving material near the orbital apocenter.
The optical emission arises from the apocenter shock, and it peaks when the most bound debris returns to the apocenter  a ``fallback time" after the star passes through pericenter and its disruption begins:\footnote{Radiation diffusion from the shocked region can lengthen the timescale.}
\begin{align}
t_{\rm fb}\simeq41{\,\rm d\,}R_{*,0}^{3/2}M_{*,0}^{-1}M_{\rm BH,6}^{1/2} \ .
	\label{eq t_fb}
\end{align}

Thus, the unbound debris is launched $\sim t_{\rm fb}$ before the optical peak.
\cite{Ryu+2020b} estimated the timescale $t_{\rm fb}=29_{-4}^{+5}\,\rm d$ within the apocenter-shock model,\footnote{They took into account the order-unity corrections due to stellar structure and relativistic effects in Eq.~\eqref{eq t_fb}.} which is consistent with our result $\simeq30\,\rm d$,
supporting identification of the outflow with the unbound debris that are launched at this time.
Additional evidence comes from comparison of the outflow speed to the expected debris speed.
Our equipartition analysis (as encapsulated in Fig.~\ref{fig energy}) suggests that the minimum outflow mass required to explain the shocked gas's energy at the last observation is $\sim 2\times 10^{-3}f_{\rm A}\, \Msun$, with $f_{\rm A}$ likely in the range $\sim 0.1 -1$ \citep{Krolik+2016,Yalinewich+2019b}.  Although most of the debris has energy corresponding to a speed at infinity $\lesssim 6000\,\rm km\,s^{-1}$ \citep{Ryu+2020}, such a small mass may be part of the high-energy tail found in the energy distribution of unbound debris for stars with mass $\gtrsim 1\, \Msun$ \citep{Ryu+2020}; we have already argued that the portion of the outflow driving the expansion we see is the small fraction of most rapidly-moving gas. The mass fraction at an energy corresponding to $13000\,f_{\rm A}^{-0.4}{\,\rm km\,s^{-1}\,}$ is close to this magnitude, particularly when the uncertainties in the equipartition analysis are taken into account.\footnote{The exponent $-0.4$ is the exponent of $f_{\rm A}$ in Eq.~\eqref{eq r_eq} for $p=2.7$.}
This idea is also consistent with the estimate by \citet{Ryu+2020b} that the disrupted stellar mass in this event was $\simeq 4\,\Msun$.  

According to the reprocessing model \citep[e.g.][]{Loeb&Ulmer1997,Strubbe&Quataert2009,Metzger&Stone2016}, an accretion disk forms when the fallback matter returns to the pericenter.\footnote{However, an  estimate of the outflow mass suggests that this model requires more mass than is typically involved in TDEs \citep{Matsumoto&Piran2021}.}
The accretion rate is larger than the Eddington rate and a massive disk wind is launched, blocking X-rays from the disk and reprocessing them to optical photons.
Such a massive disk wind can also be the origin of the radio outflow for the freely-coasting  fit.
Neither optical emission model predicts outflows launched at the optical peak, as possibly inferred from the decelerating fit.

In another model that has been suggested for this object, an outflow, possibly the result of stream self-intersections, shocks against clouds of the sort that would emit the broad emission lines of an AGN, rather than a smooth CNM \citep{Mou+2022}.
However, it should be noted that the host galaxy of AT2019dsg shows no particular AGN activity.
In addition, the density in the emission site estimated by our analysis is a factor $\sim 10^{-5\pm1}$ smaller than the density inferred for broad-line region clouds.

\subsection{The neutrino connection}

AT2019dsg coincides in position with the high-energy ($\simeq0.2\,\rm PeV$) neutrino IceCube-101001A detected on 1 October 2019 \citep[175 days after the discovery,][]{Stein+2021}.
\cite{Stein+2021} suggested that the neutrino was produced by a proton accelerated by a radio-emitting outflow whose energy is as large as $\sim10^{53}\,\rm erg$.
This is much larger than the energy estimates based on the radio observations \citep[see also][]{Cendes+2021b}.
This situation does not change even if we consider a conical outflow like a jet or the outflow-cloud interaction scenario.
\cite{Winter&Lunardini2021} proposed that the neutrino was produced by an on-axis jet which continuously injects energy to the radio-emitting site.
Similarly, their jet luminosity of $\simeq3\times10^{44}\,\rm erg\,s^{-1}$ and injected energy of $\simeq5\times10^{51}\,\rm erg$ are three orders of magnitude larger than those implied by our estimates of the energy in the emitting region.
The large discrepancy in energy could be resolved with very small equipartition parameters $\varepsilon_{\rm e}$ and $\varepsilon_{\rm B}$ are $\lesssim10^{-4}$.
However, with these energy injection rates the radio outflow would have been accelerated to  relativistic velocities, which are inconsistent with the inferred emission radii form our equipartition analysis.
Moreover, such a relativistic outflow was not seen by the VLBI observation \citep{Mohan+2021}.
\cite{Wu+2021} discussed the neutrino production from the outflow-cloud interaction, but the kinetic luminosity required in their model, $\sim10^{45}\,\rm erg\,s^{-1}$, is much larger than our estimate for what the radio emission requires (Eq.~\ref{eq luminosity}).
Clearly, scenarios in which  the neutrino production site is not related to the radio-emitting outflow \citep{Hayasaki&Yamazaki2019,Liu+2020,Murase+2020} are not constrained by our results.

\section{Summary}\label{sec summary}
TDE AT2019dsg presents rich data not only in the optical/UV and X-rays but also in the radio bands, thereby giving us a good opportunity to study the dynamics of the radio-emitting outflow.
In this work, we revisit the radio data and results of the equipartition method carried out by \cite{Stein+2021} and by \cite{Cendes+2021b}.
This method gives us the radius and energy within the radio-emitting site at each observation time when the SSA spectral peak is available.
For the equipartition radius, we fit the evolution with linear and power-law functions representing freely-coasting and decelerating outflows, respectively.
Assuming isotropic geometry, the freely-coasting fit gives an outflow velocity $\simeq13000\,\rm km\,s^{-1}$.
The deceleration fit implies $\simeq16000{\,\rm km\,s^{-1}\,}(t/38{\,\rm d})^{-0.2}$, and the inferred CNM density distribution is a little shallower than that of the former one due to the different velocity evolution.
For the most part, our results of this equipartition analysis agree with the analysis of \cite{Cendes+2021b}. 
 Important  quantitative differences  arise, however, 
in the inferred density, which we find to be smaller by a factor $\simeq 5$ due to the deep-Newtonian correction that we took into account and \cite{Cendes+2021b} ignored.

Although our fitting results largely agree with previous analysis, our interpretation is just the opposite.
\cite{Stein+2021} and \cite{Cendes+2021b} interpreted the increase of the equipartition energy  with time as due to a continuous energy injection.
We have shown here that regardless of the fit chosen, freely-coasting or deceleration, continuous energy injection requires some sort of fine tuning.
The CNM density profile must be matched to the rate of energy injection in order to keep the velocity almost constant, the energy injection rate of $\dot{E}_{\rm kin}\propto t^{-0.35}$ varies with time in a way much different from the mass fallback rate to the BH, and the total injected energy is very small compared to the overall energy budget of the system.
Moreover, injection of energy by a jet would produce a very powerful reverse shock that is not seen.

On the other hand, the increase in energy can be naturally interpreted as being drawn from the kinetic energy of an instantaneously ejected outflow.
In particular, this outflow could be the unbound debris ejected during the stellar disruption \citep[as suggested for ASASSN-14li,][]{Krolik+2016,Yalinewich+2019b}.
The outflow's total kinetic energy is much larger than that of the shocked region; hence, the velocity automatically remains essentially constant.
Importantly, this is a natural outcome of an early stage of the outflow-CNM interaction very closely analogous to the case of supernova remnants.
As the outflow sweeps up more and more matter, the energy within the radio-emitting region increases.

To summarize, the basic picture in which radio emission is produced by a shock running into external material naturally leads to continuous energy transfer from material driving the shock into the shocked region.
Thus, in any shock-driven system in which the cooling time is longer than the dynamical time, and in particular here and in other radio TDEs, (e.g.   ASASSN-14li and CNSS J0019+00) this is the simplest and most natural explanation for the gradual growth of the energy in the radio emitting region.

\section*{acknowledgments}
We thank Chi-Ho Chan, Assaf Horesh, Christopher Irwin, and Ehud Nakar for fruitful discussions and helpful comments.
T.M. thanks the Yukawa Institute for Theoretical Physics at Kyoto University. Discussions during the YITP workshop YITP-T-21-05 on ``Extreme Outflows in Astrophysical Transients'' were useful for this work.
This work is supported in part by JSPS Postdoctral Fellowship, Kakenhi No. 19J00214 (T.M.) and by ERC advanced grant ``TReX'' (T.P.).

\section*{data availability}
The data underlying this article will be shared on reasonable request to the corresponding author.

\bibliographystyle{mnras}
\bibliography{reference_matsumoto}

\begin{thebibliography}{}
\makeatletter
\relax
\def\mn@urlcharsother{\let\do\@makeother \do\$\do\&\do\#\do\^\do\_\do\%\do\~}
\def\mn@doi{\begingroup\mn@urlcharsother \@ifnextchar [ {\mn@doi@}
  {\mn@doi@[]}}
\def\mn@doi@[#1]#2{\def\@tempa{#1}\ifx\@tempa\@empty \href
  {http://dx.doi.org/#2} {doi:#2}\else \href {http://dx.doi.org/#2} {#1}\fi
  \endgroup}
\def\mn@eprint#1#2{\mn@eprint@#1:#2::\@nil}
\def\mn@eprint@arXiv#1{\href {http://arxiv.org/abs/#1} {{\tt arXiv:#1}}}
\def\mn@eprint@dblp#1{\href {http://dblp.uni-trier.de/rec/bibtex/#1.xml}
  {dblp:#1}}
\def\mn@eprint@#1:#2:#3:#4\@nil{\def\@tempa {#1}\def\@tempb {#2}\def\@tempc
  {#3}\ifx \@tempc \@empty \let \@tempc \@tempb \let \@tempb \@tempa \fi \ifx
  \@tempb \@empty \def\@tempb {arXiv}\fi \@ifundefined
  {mn@eprint@\@tempb}{\@tempb:\@tempc}{\expandafter \expandafter \csname
  mn@eprint@\@tempb\endcsname \expandafter{\@tempc}}}

\bibitem[\protect\citeauthoryear{{Alexander}, {Berger}, {Guillochon},
  {Zauderer}  \& {Williams}}{{Alexander} et~al.}{2016}]{Alexander+2016}
{Alexander} K.~D.,  {Berger} E.,  {Guillochon} J.,  {Zauderer} B.~A.,
  {Williams} P.~K.~G.,  2016, \mn@doi [\apjl] {10.3847/2041-8205/819/2/L25},
  \href {https://ui.adsabs.harvard.edu/abs/2016ApJ...819L..25A} {819, L25}

\bibitem[\protect\citeauthoryear{{Alexander}, {van Velzen}, {Horesh}  \&
  {Zauderer}}{{Alexander} et~al.}{2020}]{Alexander+2020}
{Alexander} K.~D.,  {van Velzen} S.,  {Horesh} A.,   {Zauderer} B.~A.,  2020,
  \mn@doi [\ssr] {10.1007/s11214-020-00702-w}, \href
  {https://ui.adsabs.harvard.edu/abs/2020SSRv..216...81A} {216, 81}

\bibitem[\protect\citeauthoryear{{Anderson} et~al.,}{{Anderson}
  et~al.}{2020}]{Anderson+2020}
{Anderson} M.~M.,  et~al., 2020, \mn@doi [\apj] {10.3847/1538-4357/abb94b},
  \href {https://ui.adsabs.harvard.edu/abs/2020ApJ...903..116A} {903, 116}

\bibitem[\protect\citeauthoryear{{Baganoff} et~al.,}{{Baganoff}
  et~al.}{2003}]{Baganoff+2003}
{Baganoff} F.~K.,  et~al., 2003, \mn@doi [\apj] {10.1086/375145}, \href
  {http://ads.nao.ac.jp/abs/2003ApJ...591..891B} {591, 891}

\bibitem[\protect\citeauthoryear{{Barniol Duran}, {Nakar}  \& {Piran}}{{Barniol
  Duran} et~al.}{2013}]{BarniolDuran+2013}
{Barniol Duran} R.,  {Nakar} E.,   {Piran} T.,  2013, \mn@doi [\apj]
  {10.1088/0004-637X/772/1/78}, \href
  {https://ui.adsabs.harvard.edu/abs/2013ApJ...772...78B} {772, 78}

\bibitem[\protect\citeauthoryear{{Brown}, {Holoien}, {Auchettl}, {Stanek},
  {Kochanek}, {Shappee}, {Prieto}  \& {Grupe}}{{Brown}
  et~al.}{2017}]{Brown+2017}
{Brown} J.~S.,  {Holoien} T.~W.~S.,  {Auchettl} K.,  {Stanek} K.~Z.,
  {Kochanek} C.~S.,  {Shappee} B.~J.,  {Prieto} J.~L.,   {Grupe} D.,  2017,
  \mn@doi [\mnras] {10.1093/mnras/stx033}, \href
  {https://ui.adsabs.harvard.edu/abs/2017MNRAS.466.4904B} {466, 4904}

\bibitem[\protect\citeauthoryear{{Cannizzaro} et~al.,}{{Cannizzaro}
  et~al.}{2021}]{Cannizzaro+2021}
{Cannizzaro} G.,  et~al., 2021, \mn@doi [\mnras] {10.1093/mnras/stab851}, \href
  {https://ui.adsabs.harvard.edu/abs/2021MNRAS.504..792C} {504, 792}

\bibitem[\protect\citeauthoryear{{Cendes}, {Alexander}, {Berger}, {Eftekhari},
  {Williams}  \& {Chornock}}{{Cendes} et~al.}{2021}]{Cendes+2021b}
{Cendes} Y.,  {Alexander} K.~D.,  {Berger} E.,  {Eftekhari} T.,  {Williams}
  P.~K.~G.,   {Chornock} R.,  2021, \mn@doi [\apj] {10.3847/1538-4357/ac110a},
  \href {https://ui.adsabs.harvard.edu/abs/2021ApJ...919..127C} {919, 127}

\bibitem[\protect\citeauthoryear{{Cenko} et~al.,}{{Cenko}
  et~al.}{2016}]{Cenko+2016}
{Cenko} S.~B.,  et~al., 2016, \mn@doi [\apjl] {10.3847/2041-8205/818/2/L32},
  \href {https://ui.adsabs.harvard.edu/abs/2016ApJ...818L..32C} {818, L32}

\bibitem[\protect\citeauthoryear{{Chevalier}}{{Chevalier}}{1998}]{Chevalier1998}
{Chevalier} R.~A.,  1998, \mn@doi [\apj] {10.1086/305676}, \href
  {http://adsabs.harvard.edu/abs/1998ApJ...499..810C} {499, 810}

\bibitem[\protect\citeauthoryear{{Dai}, {McKinney}, {Roth}, {Ramirez-Ruiz}  \&
  {Miller}}{{Dai} et~al.}{2018}]{Dai+2018}
{Dai} L.,  {McKinney} J.~C.,  {Roth} N.,  {Ramirez-Ruiz} E.,   {Miller} M.~C.,
  2018, \mn@doi [\apjl] {10.3847/2041-8213/aab429}, \href
  {https://ui.adsabs.harvard.edu/abs/2018ApJ...859L..20D} {859, L20}

\bibitem[\protect\citeauthoryear{{Gillessen} et~al.,}{{Gillessen}
  et~al.}{2019}]{Gillessen+2019}
{Gillessen} S.,  et~al., 2019, \mn@doi [\apj] {10.3847/1538-4357/aaf4f8}, \href
  {https://ui.adsabs.harvard.edu/abs/2019ApJ...871..126G} {871, 126}

\bibitem[\protect\citeauthoryear{{Hayasaki} \& {Yamazaki}}{{Hayasaki} \&
  {Yamazaki}}{2019}]{Hayasaki&Yamazaki2019}
{Hayasaki} K.,  {Yamazaki} R.,  2019, \mn@doi [\apj]
  {10.3847/1538-4357/ab44ca}, \href
  {https://ui.adsabs.harvard.edu/abs/2019ApJ...886..114H} {886, 114}

\bibitem[\protect\citeauthoryear{{Hills}}{{Hills}}{1975}]{Hills1975}
{Hills} J.~G.,  1975, \mn@doi [\nat] {10.1038/254295a0}, \href
  {https://ui.adsabs.harvard.edu/abs/1975Natur.254..295H} {254, 295}

\bibitem[\protect\citeauthoryear{{Holoien} et~al.,}{{Holoien}
  et~al.}{2016}]{Holoien+2016}
{Holoien} T.~W.~S.,  et~al., 2016, \mn@doi [\mnras] {10.1093/mnras/stv2486},
  \href {https://ui.adsabs.harvard.edu/abs/2016MNRAS.455.2918H} {455, 2918}

\bibitem[\protect\citeauthoryear{{Huang} \& {Cheng}}{{Huang} \&
  {Cheng}}{2003}]{Huang&Cheng2003}
{Huang} Y.~F.,  {Cheng} K.~S.,  2003, \mn@doi [\mnras]
  {10.1046/j.1365-8711.2003.06430.x}, \href
  {https://ui.adsabs.harvard.edu/abs/2003MNRAS.341..263H} {341, 263}

\bibitem[\protect\citeauthoryear{{Jiang}, {Dou}, {Wang}, {Yang}, {Lyu}  \&
  {Zhou}}{{Jiang} et~al.}{2016}]{Jiang+2016}
{Jiang} N.,  {Dou} L.,  {Wang} T.,  {Yang} C.,  {Lyu} J.,   {Zhou} H.,  2016,
  \mn@doi [\apjl] {10.3847/2041-8205/828/1/L14}, \href
  {https://ui.adsabs.harvard.edu/abs/2016ApJ...828L..14J} {828, L14}

\bibitem[\protect\citeauthoryear{{Krolik}, {Piran}, {Svirski}  \&
  {Cheng}}{{Krolik} et~al.}{2016}]{Krolik+2016}
{Krolik} J.,  {Piran} T.,  {Svirski} G.,   {Cheng} R.~M.,  2016, \mn@doi [\apj]
  {10.3847/0004-637X/827/2/127}, \href
  {https://ui.adsabs.harvard.edu/abs/2016ApJ...827..127K} {827, 127}

\bibitem[\protect\citeauthoryear{{Landau} \& {Lifshitz}}{{Landau} \&
  {Lifshitz}}{1987}]{Landau&Lifshitz1987}
{Landau} L.~D.,  {Lifshitz} E.~M.,  1987, {Fluid Mechanics}

\bibitem[\protect\citeauthoryear{{Liu}, {Xi}  \& {Wang}}{{Liu}
  et~al.}{2020}]{Liu+2020}
{Liu} R.-Y.,  {Xi} S.-Q.,   {Wang} X.-Y.,  2020, \mn@doi [\prd]
  {10.1103/PhysRevD.102.083028}, \href
  {https://ui.adsabs.harvard.edu/abs/2020PhRvD.102h3028L} {102, 083028}

\bibitem[\protect\citeauthoryear{{Loeb} \& {Ulmer}}{{Loeb} \&
  {Ulmer}}{1997}]{Loeb&Ulmer1997}
{Loeb} A.,  {Ulmer} A.,  1997, \mn@doi [\apj] {10.1086/304814}, \href
  {https://ui.adsabs.harvard.edu/abs/1997ApJ...489..573L} {489, 573}

\bibitem[\protect\citeauthoryear{{Lu} \& {Bonnerot}}{{Lu} \&
  {Bonnerot}}{2020}]{Lu&Bonnerot2020}
{Lu} W.,  {Bonnerot} C.,  2020, \mn@doi [\mnras] {10.1093/mnras/stz3405}, \href
  {https://ui.adsabs.harvard.edu/abs/2020MNRAS.492..686L} {492, 686}

\bibitem[\protect\citeauthoryear{{Matsumoto} \& {Piran}}{{Matsumoto} \&
  {Piran}}{2021a}]{Matsumoto&Piran2021}
{Matsumoto} T.,  {Piran} T.,  2021a, \mn@doi [\mnras] {10.1093/mnras/stab240},
  \href {https://ui.adsabs.harvard.edu/abs/2021MNRAS.502.3385M} {502, 3385}

\bibitem[\protect\citeauthoryear{{Matsumoto} \& {Piran}}{{Matsumoto} \&
  {Piran}}{2021b}]{Matsumoto&Piran2021b}
{Matsumoto} T.,  {Piran} T.,  2021b, \mn@doi [\mnras] {10.1093/mnras/stab2418},
  \href {https://ui.adsabs.harvard.edu/abs/2021MNRAS.507.4196M} {507, 4196}

\bibitem[\protect\citeauthoryear{{Metzger} \& {Stone}}{{Metzger} \&
  {Stone}}{2016}]{Metzger&Stone2016}
{Metzger} B.~D.,  {Stone} N.~C.,  2016, \mn@doi [\mnras]
  {10.1093/mnras/stw1394}, \href
  {https://ui.adsabs.harvard.edu/abs/2016MNRAS.461..948M} {461, 948}

\bibitem[\protect\citeauthoryear{{Miller} et~al.,}{{Miller}
  et~al.}{2015}]{Miller+2015}
{Miller} J.~M.,  et~al., 2015, \mn@doi [\nat] {10.1038/nature15708}, \href
  {https://ui.adsabs.harvard.edu/abs/2015Natur.526..542M} {526, 542}

\bibitem[\protect\citeauthoryear{{Mohan}, {An}, {Zhang}, {Yang}, {Yang}  \&
  {Wang}}{{Mohan} et~al.}{2021}]{Mohan+2021}
{Mohan} P.,  {An} T.,  {Zhang} Y.,  {Yang} J.,  {Yang} X.,   {Wang} A.,  2021,
  arXiv e-prints, \href {https://ui.adsabs.harvard.edu/abs/2021arXiv210615799M}
  {p. arXiv:2106.15799}

\bibitem[\protect\citeauthoryear{{Mou}, {Wang}, {Wang}  \& {Yang}}{{Mou}
  et~al.}{2022}]{Mou+2022}
{Mou} G.,  {Wang} T.,  {Wang} W.,   {Yang} J.,  2022, \mn@doi [\mnras]
  {10.1093/mnras/stab3742}, \href
  {https://ui.adsabs.harvard.edu/abs/2022MNRAS.510.3650M} {510, 3650}

\bibitem[\protect\citeauthoryear{{Murase}, {Kimura}, {Zhang}, {Oikonomou}  \&
  {Petropoulou}}{{Murase} et~al.}{2020}]{Murase+2020}
{Murase} K.,  {Kimura} S.~S.,  {Zhang} B.~T.,  {Oikonomou} F.,   {Petropoulou}
  M.,  2020, \mn@doi [\apj] {10.3847/1538-4357/abb3c0}, \href
  {https://ui.adsabs.harvard.edu/abs/2020ApJ...902..108M} {902, 108}

\bibitem[\protect\citeauthoryear{{Pacholczyk}}{{Pacholczyk}}{1970}]{Pacholczyk1970}
{Pacholczyk} A.~G.,  1970, {Radio astrophysics. Nonthermal processes in
  galactic and extragalactic sources}

\bibitem[\protect\citeauthoryear{{Phinney}}{{Phinney}}{1989}]{Phinney1989}
{Phinney} E.~S.,  1989, in {Morris} M.,  ed.,  IAU Symposium Vol. 136, The
  Center of the Galaxy. p.~543

\bibitem[\protect\citeauthoryear{{Piran}, {Svirski}, {Krolik}, {Cheng}  \&
  {Shiokawa}}{{Piran} et~al.}{2015}]{Piran+2015}
{Piran} T.,  {Svirski} G.,  {Krolik} J.,  {Cheng} R.~M.,   {Shiokawa} H.,
  2015, \mn@doi [\apj] {10.1088/0004-637X/806/2/164}, \href
  {http://adsabs.harvard.edu/abs/2015ApJ...806..164P} {806, 164}

\bibitem[\protect\citeauthoryear{{Rees}}{{Rees}}{1988}]{Rees1988}
{Rees} M.~J.,  1988, \mn@doi [\nat] {10.1038/333523a0}, \href
  {https://ui.adsabs.harvard.edu/abs/1988Natur.333..523R} {333, 523}

\bibitem[\protect\citeauthoryear{{Roth}, {Rossi}, {Krolik}, {Piran}, {Mockler}
  \& {Kasen}}{{Roth} et~al.}{2020}]{Roth+2020}
{Roth} N.,  {Rossi} E.~M.,  {Krolik} J.,  {Piran} T.,  {Mockler} B.,   {Kasen}
  D.,  2020, \mn@doi [\ssr] {10.1007/s11214-020-00735-1}, \href
  {https://ui.adsabs.harvard.edu/abs/2020SSRv..216..114R} {216, 114}

\bibitem[\protect\citeauthoryear{{Ryu}, {Krolik}  \& {Piran}}{{Ryu}
  et~al.}{2020a}]{Ryu+2020b}
{Ryu} T.,  {Krolik} J.,   {Piran} T.,  2020a, \mn@doi [\apj]
  {10.3847/1538-4357/abbf4d}, \href
  {https://ui.adsabs.harvard.edu/abs/2020ApJ...904...73R} {904, 73}

\bibitem[\protect\citeauthoryear{{Ryu}, {Krolik}, {Piran}  \& {Noble}}{{Ryu}
  et~al.}{2020b}]{Ryu+2020}
{Ryu} T.,  {Krolik} J.,  {Piran} T.,   {Noble} S.~C.,  2020b, \mn@doi [\apj]
  {10.3847/1538-4357/abb3cd}, \href
  {https://ui.adsabs.harvard.edu/abs/2020ApJ...904...99R} {904, 99}

\bibitem[\protect\citeauthoryear{{Sari}, {Piran}  \& {Narayan}}{{Sari}
  et~al.}{1998}]{Sari+1998}
{Sari} R.,  {Piran} T.,   {Narayan} R.,  1998, \mn@doi [\apjl]
  {10.1086/311269}, \href {http://adsabs.harvard.edu/abs/1998ApJ...497L..17S}
  {497, L17}

\bibitem[\protect\citeauthoryear{{Saxton}, {Komossa}, {Auchettl}  \&
  {Jonker}}{{Saxton} et~al.}{2020}]{Saxton+2020}
{Saxton} R.,  {Komossa} S.,  {Auchettl} K.,   {Jonker} P.~G.,  2020, \mn@doi
  [\ssr] {10.1007/s11214-020-00708-4}, \href
  {https://ui.adsabs.harvard.edu/abs/2020SSRv..216...85S} {216, 85}

\bibitem[\protect\citeauthoryear{{Scott} \& {Readhead}}{{Scott} \&
  {Readhead}}{1977}]{Scott&Readhead1977}
{Scott} M.~A.,  {Readhead} A.~C.~S.,  1977, \mn@doi [\mnras]
  {10.1093/mnras/180.4.539}, \href
  {https://ui.adsabs.harvard.edu/abs/1977MNRAS.180..539S} {180, 539}

\bibitem[\protect\citeauthoryear{{Sironi} \& {Giannios}}{{Sironi} \&
  {Giannios}}{2013}]{Sironi&Giannios2013}
{Sironi} L.,  {Giannios} D.,  2013, \mn@doi [\apj]
  {10.1088/0004-637X/778/2/107}, \href
  {https://ui.adsabs.harvard.edu/abs/2013ApJ...778..107S} {778, 107}

\bibitem[\protect\citeauthoryear{{S{\k{a}}dowski}, {Tejeda}, {Gafton},
  {Rosswog}  \& {Abarca}}{{S{\k{a}}dowski} et~al.}{2016}]{Sadowski+2016}
{S{\k{a}}dowski} A.,  {Tejeda} E.,  {Gafton} E.,  {Rosswog} S.,   {Abarca} D.,
  2016, \mn@doi [\mnras] {10.1093/mnras/stw589}, \href
  {https://ui.adsabs.harvard.edu/abs/2016MNRAS.458.4250S} {458, 4250}

\bibitem[\protect\citeauthoryear{{Stein} et~al.,}{{Stein}
  et~al.}{2021}]{Stein+2021}
{Stein} R.,  et~al., 2021, \mn@doi [Nature Astronomy]
  {10.1038/s41550-020-01295-8}, \href
  {https://ui.adsabs.harvard.edu/abs/2021NatAs...5..510S} {5, 510}

\bibitem[\protect\citeauthoryear{{Stone}, {Vasiliev}, {Kesden}, {Rossi},
  {Perets}  \& {Amaro-Seoane}}{{Stone} et~al.}{2020}]{Stone+2020}
{Stone} N.~C.,  {Vasiliev} E.,  {Kesden} M.,  {Rossi} E.~M.,  {Perets} H.~B.,
  {Amaro-Seoane} P.,  2020, \mn@doi [\ssr] {10.1007/s11214-020-00651-4}, \href
  {https://ui.adsabs.harvard.edu/abs/2020SSRv..216...35S} {216, 35}

\bibitem[\protect\citeauthoryear{{Strubbe} \& {Quataert}}{{Strubbe} \&
  {Quataert}}{2009}]{Strubbe&Quataert2009}
{Strubbe} L.~E.,  {Quataert} E.,  2009, \mn@doi [\mnras]
  {10.1111/j.1365-2966.2009.15599.x}, \href
  {https://ui.adsabs.harvard.edu/abs/2009MNRAS.400.2070S} {400, 2070}

\bibitem[\protect\citeauthoryear{{Winter} \& {Lunardini}}{{Winter} \&
  {Lunardini}}{2021}]{Winter&Lunardini2021}
{Winter} W.,  {Lunardini} C.,  2021, \mn@doi [Nature Astronomy]
  {10.1038/s41550-021-01305-3}, \href
  {https://ui.adsabs.harvard.edu/abs/2021NatAs...5..472W} {5, 472}

\bibitem[\protect\citeauthoryear{{Wu}, {Mou}, {Wang}, {Wang}  \& {Li}}{{Wu}
  et~al.}{2021}]{Wu+2021}
{Wu} H.-J.,  {Mou} G.-B.,  {Wang} K.,  {Wang} W.,   {Li} Z.,  2021, arXiv
  e-prints, \href {https://ui.adsabs.harvard.edu/abs/2021arXiv211201748W} {p.
  arXiv:2112.01748}

\bibitem[\protect\citeauthoryear{{Yalinewich}, {Steinberg}, {Piran}  \&
  {Krolik}}{{Yalinewich} et~al.}{2019}]{Yalinewich+2019b}
{Yalinewich} A.,  {Steinberg} E.,  {Piran} T.,   {Krolik} J.~H.,  2019, \mn@doi
  [\mnras] {10.1093/mnras/stz1567}, \href
  {https://ui.adsabs.harvard.edu/abs/2019MNRAS.487.4083Y} {487, 4083}

\bibitem[\protect\citeauthoryear{{van Velzen} et~al.,}{{van Velzen}
  et~al.}{2016}]{vanVelzen+2016}
{van Velzen} S.,  et~al., 2016, \mn@doi [Science] {10.1126/science.aad1182},
  \href {https://ui.adsabs.harvard.edu/abs/2016Sci...351...62V} {351, 62}

\bibitem[\protect\citeauthoryear{{van Velzen}, {Holoien}, {Onori}, {Hung}  \&
  {Arcavi}}{{van Velzen} et~al.}{2020}]{vanVelzen+2020}
{van Velzen} S.,  {Holoien} T. W.~S.,  {Onori} F.,  {Hung} T.,   {Arcavi} I.,
  2020, \mn@doi [\ssr] {10.1007/s11214-020-00753-z}, \href
  {https://ui.adsabs.harvard.edu/abs/2020SSRv..216..124V} {216, 124}

\bibitem[\protect\citeauthoryear{{van Velzen} et~al.,}{{van Velzen}
  et~al.}{2021}]{vanVelzen+2021}
{van Velzen} S.,  et~al., 2021, \mn@doi [\apj] {10.3847/1538-4357/abc258},
  \href {https://ui.adsabs.harvard.edu/abs/2021ApJ...908....4V} {908, 4}

\makeatother
\end{thebibliography}

\label{lastpage}
\end{document}